\documentclass[aip,rsi,reprint,superscriptaddress]{revtex4-1}
\usepackage{graphicx}
\usepackage{tikz}   
\usepackage{float}
\usepackage{overpic}
\usepackage{todonotes}
\usepackage[colorlinks=true,linkcolor=blue,citecolor=blue,urlcolor=blue]{hyperref}

\begin{document}


\title{Versatile Cold Atom Source for Multi-Species Experiments}



\author{A. Paris-Mandoki}
\affiliation{School of Physics and Astronomy, University of Nottingham, UK}
\author{M. D. Jones}
\affiliation{School of Physics and Astronomy, University of Nottingham, UK}
\author{J. Nute}
\affiliation{School of Physics and Astronomy, University of Nottingham, UK}
\author{J. Wu}
\affiliation{School of Physics and Astronomy, University of Nottingham, UK}
\affiliation{State Key Laboratory of Quantum Optics and Quantum Optics Devices, Institute of Laser Spectroscopy, College of
Physics and Electronics Engineering, Shanxi University, Taiyuan, 030006, China}
\author{S. Warriar}
\affiliation{School of Physics and Astronomy, University of Nottingham, UK}
\author{L. Hackerm{\"u}ller}
\email[]{lucia.hackermuller@nottingham.ac.uk}
\affiliation{School of Physics and Astronomy, University of Nottingham, UK}

%


\date{\today}

\begin{abstract}
We present a dual-species oven and Zeeman slower setup capable of producing slow, high-flux atomic beams for loading magneto-optical traps. Our compact and versatile system is based on electronic switching between different magnetic field profiles and is applicable to a wide range of multi-species experiments. We give details of the vacuum setup, coils and simple electronic circuitry. In addition, we demonstrate the performance of our system by optimized, sequential loading of magneto-optical traps of lithium-6 and cesium-133.
\end{abstract}


\maketitle


\section{\label{Introduction}Introduction}

Experiments employing multiple atomic species are becoming increasingly important in the field of ultracold atomic physics. Diverse applications and fundamental questions arise as a result of the interactions between species, which may be bosonic or fermionic \cite{Truscott2001, Hadzibabic2002, Roati2002, Goldwin2004} and have different masses \cite{Jag2014}, spins or scattering properties. Pairing of two species to create heteronuclear molecules is carried out in experiments motivated by exploitation of the long range dipole interaction \cite{Ni2008}. In addition multiple species have also been used as a tool to produce degenerate gases through sympathetic evaporative cooling \cite{Modugno2001, Mudrich2002}. 

Loading a magneto-optical trap (MOT) from an atomic source is a ubiquitous requirement of cold atomic experiments. For this task a high-flux beam of atoms traveling at velocities within the capture range of the MOT, from a source remote to the MOT chamber, is ideal. A remote atom source facilitates lower pressures in the MOT chamber, which allows longer lifetimes of ultracold samples without the complication and bulk of an additional chamber and transport scheme. These conditions are commonly met by use of a Zeeman slower, first demonstrated by Phillips and Metcalf \cite{Phillips1982} or alternatively by a two-dimensional MOT (2DMOT). The first of these has several advantages in complex experimental setups. Zeeman slowers require less laser power, fewer optical components and are less sensitive to alignment. The use of only a single vacuum tube makes them suitable for gases which are aggressive to glass. In addition, the generally lower densities than a 2DMOT lead to reduced collisional atom losses due to interspecies interactions\cite{Schloder1999} in multiple species experiments.

In these experiments the use of a Zeeman slower is therefore highly attractive, however, different species benefit from slowers with different magnetic field profiles. One solution is to use separate Zeeman slowers for each species, however, the repercussion of this is a larger vacuum apparatus and loss of optical access to the MOT. It is therefore highly advantageous to use a multi-species oven and single slower.

Different approaches to the design of a multi-species Zeeman slower have been developed in the past. A dual-species static magnetic field profile slower\cite{Marti2010} requires compromise in its capability for each element. An alternative approach is to employ an array of servo motors to controllably position permanent magnets to produce different magnetic field profiles\cite{Reinaudi2012}. However, this results in a bulky arrangement with a large number of mechanical parts and long switching times between different configurations.

The slower presented in this work is suitable for consecutive MOT-loading of two different species by changing the magnetic field profile accordingly. We use an array of coil sections to generate the magnetic field and by means of simple electronics we switch between two magnetic field profiles by tailoring the current in different sections of the slower. The resultant atom beams sequentially load two MOTs. By holding the atoms of the first MOT in an optical dipole trap, both species can be combined after loading the second MOT\cite{Tung2013}.

The design is fast-switching, compact and is implementable \textit{in situ} with existing slowers which have been designed for use with a single species. We give details of the design of the magnetic field profiles in Section~\ref{Design and Calculations} and of the experimental apparatus and electronics in Section~\ref{Experimental Realization}. In addition we discuss variations of this versatile design to customize it for experiments in which an increasing-field or spin-flip profile may be preferred. In Section~\ref{Measurements and Discussion} we demonstrate optimization of MOT loading for lithium-6 and cesium-133. This is a combination with large mass imbalance and thus is a rigorous test case for the system.

\section{\label{Design and Calculations}Design and Calculation}

A Zeeman slower design should conform to the following specifications. A large proportion of the atoms from the oven source must be slowed to a velocity which can be captured by the MOT. The light for the slower necessarily passes through the MOT, however, it should be sufficiently detuned such that it has negligible effect on it. Finally, the slower should maintain maximum achievable deceleration of the atoms at all points along the slower, such that it is short without compromising in capture velocity. A shorter slower loses fewer atoms due to divergence of the beam and results in a more compact setup.

Zeeman slowers fall into three categories based on the shape of their magnetic field profile. An increasing-field slower requires the highest laser detuning from resonance; atoms in the MOT are negligibly affected by the slower light. On the other hand, strong magnetic fields from the slower distort the MOT's quadrupole field. A spin-flip slower, where the magnetic field has a change of sign through the slower, has the advantages of moderate magnetic field disturbance whilst retaining a detuning where the atoms in the MOT are not affected. This arrangement has the drawback that in the vicinity of the zero of the field within the slower, the atomic transition is not closed and many atoms fall out of the cooling cycle. Lastly, with a decreasing-field slower, the high magnetic fields are spatially well separated from the MOT and detunings larger than 15 linewidths can still be used.

The first step in the design process of a slower is to calculate the magnetic field profile which results in the maximum reduction in velocity in the shortest slower. As the scattering rate cannot exceed half of the atomic transition linewidth ${\Gamma}$, the maximum achievable deceleration is
\begin{equation}
\label{eq:amax} a_{\mathrm{max}} = {{\Gamma} \over {2}} {{\hbar k} \over {m}},
\end{equation}
where $\hbar$ is the reduced Planck constant, $k$ is the wavenumber of the light corresponding to the atomic transition and $m$ is the atomic mass. In practice, due to imperfect slower field and limited laser power only a fraction of this can be achieved in the laboratory. The fraction is denoted by $\eta$ and relates $a_{\mathrm{max}}$ with the achievable acceleration $a$ by
\begin{equation}
\label{} a = \eta a_{\mathrm{max}}.
\end{equation}
It is typical\citep{Foot2005} for a slower operate at $\eta > 0.5$, we therefore use this value during the design process. The magnetic field profile to achieve deceleration $a$ at every position $z$ along its length is given by
\begin{equation}
\label{eq:1} B(z) = {{\hbar} \over {\mu}}\left( {\delta + k \sqrt{{v_i}^2 -2az}} \right),
\end{equation}
where $\mu$ is the magnetic moment of the transition, $\delta$ is the detuning from the atomic transition,  and $v_i$ is the maximum capture velocity of the slower. The capture velocity is determined by the length $L$ and final velocity $v_f$ of the slower;
\begin{equation}
\label{eq:2} v_i = \sqrt{{v_f}^2 + 2aL}.
\end{equation}
The final velocity value must be chosen such that it lies within the capture velocity of the MOT.
Designing an optimal slower for two different species would result in different lengths $L$ for each species. By modifying the initial velocity $v_i$ as well as the acceleration $a$ in an interval, where the total resulting efficiency is still acceptable, we achieve a slower design which is suitable for two different species.  

For the specific case of our system the two species of interest are lithium and cesium. These have a particularly large difference in mass and initial velocity which makes this combination a good test case for the system. The design parameters are given by the following conditions. To produce a high flux from an effusive lithium oven, a vapor pressure of $10^{-4}$~mbar is required. This is achieved by heating the oven to 670~K, which yields a most probable atomic velocity of 1360~m/s from the Maxwell-Boltzmann distribution \cite{Lide2013}. We chose to red-detune the slower light by 120~MHz from the atomic resonance in order to avoid significant effect of the light on the atoms in the MOT. Using Equations~\ref{eq:amax}~to~\ref{eq:2} with $v_i=1360$~m/s and setting $v_f=200$~m/s, the capture velocity of the MOT, would result in an optimum slower length of 1.15~m  and magnetic fields in excess of 1300~G. Instead, we chose $v_i=900$~m/s which 
still encompasses a sufficiently large portion of the high initial flux. This allows the design of a 0.4~m long slower with a maximum magnetic field of 800~G.


To provide a source of cesium, an oven at a temperature of 370~K was designed. This temperature corresponds to a most probable initial velocity of 215~m/s, to be reduced by the slower to a MOT capture velocity of 40~m/s. Using slower light red detuned from resonance by 95~MHz required that the slower fields are around an order of magnitude lower than those for lithium. With a slower length of 0.4~m atoms with $v_i=215$~m/s can be captured in the MOT. 

\begin{figure}
\includegraphics[width=0.45\textwidth]{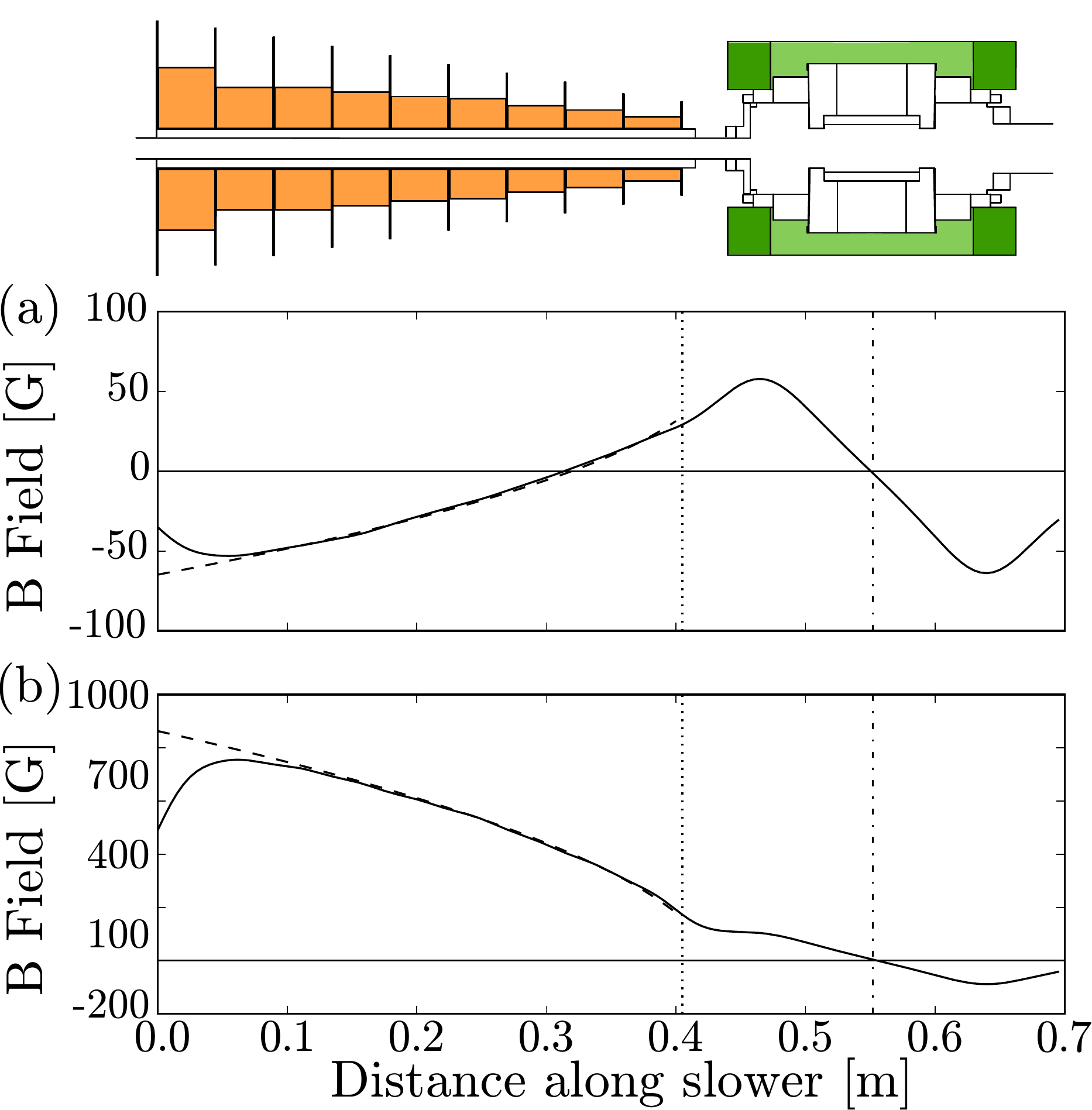}%
\caption{\label{calc_profiles_design} Calculated magnetic field profiles of the Zeeman slower for (a) cesium and (b) lithium. Dashed lines are the expected ideal magnetic field profile to achieve constant deceleration using $\eta=0.5$. Solid lines result from a model of nine coil sections where the number of layers is varied in each section to match the ideal profile for a fixed current for lithium. For cesium the same is achieved by fixing the number of layers and varying the current in each coil. The vertical dotted line indicates the end of the slower coils at 0.40~m. The vertical dash-dotted line indicates the central position of the MOT. A cross section of the experimental apparatus, to scale with the horizontal plot axes, is shown with the magnetic coils highlighted.}
\end{figure}

The resulting calculated magnetic field profiles for constant deceleration are shown in Figure~\ref{calc_profiles_design}. To find the actual coil configuration which would create this field we assume a system of nine coil sections, which is sufficient to form a smooth magnetic field profile. The slower field on the axis of the atom beam was then calculated by summing the field of infinitely thin wire loops at the center position of each winding of each coil of the slower. In the same way, the field from coils used to produce the MOT was also included since this has a strong contribution at one end of the slower. 




For lithium, a current of 11~A was chosen to be used throughout the slower and the calculated magnetic field profile was then matched to the ideal curve by varying the number of layers of windings in each coil in the model. The cesium profile would then be produced by changing the current in each of the coils.

Figure~\ref{calc_profiles_design} shows a comparison of the calculated optimal magnetic field profiles (broken line) for the two species with the field created by the nine coil sections (full line). For lithium the profile is a decreasing-field slower while the cesium profile passes through zero field. 


\section{\label{Experimental Realization}Experimental Realization}

\subsection{Overview}

\begin{figure}
\includegraphics[width=0.45\textwidth,grid]{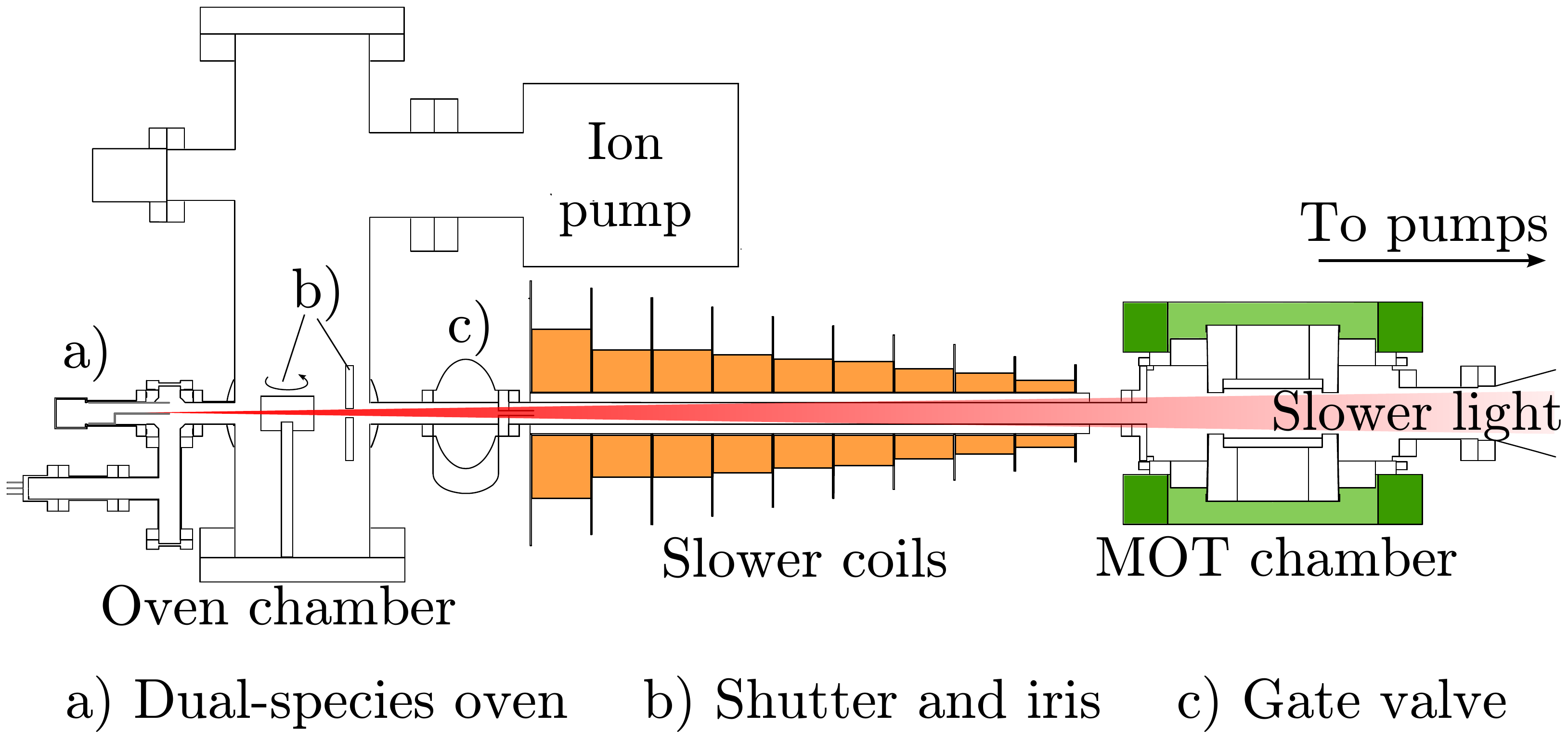}
\caption{\label{apparatus_profile} Cross-section of the experimental apparatus. Shaded regions show the coils used for slowing and trapping the atoms. Also shown is the convergent slowing light.}
\end{figure}

The context of the atom source and slower in the full experimental apparatus is shown in Figure~\ref{apparatus_profile}. The atom source is a dual-species oven on the far left of the figure, which is described in detail in the following section. After leaving the oven the atoms form a beam which is collimated by two apertures: an iris in the oven chamber and a differential pumping tube at the beginning of the actual slower. A rotating shutter can be used to block the flow of both species into the slower once a MOT has been loaded.

The oven chamber and the Zeeman slower are connected through a gate valve and a 3~mm diameter, 40~mm long differential pumping tube. The tube allows two ion pumps and titanium sublimation pumping to maintain a pressure difference between the oven chamber at $10^{-9}$~mbar and the chamber for magneto-optical trapping (MOT chamber) at $2\times 10^{-11}$~mbar. The gate valve allows cesium dispensers and lithium to be replaced without loss of vacuum in the MOT chamber.

The performance of our two-species source and slower was tested by loading a two-species magneto-optical trap in this setup.

\subsection{Dual-species oven\label{oven}}

The main design goals for our two species oven were simplicity and a comparable high atom flux for both species. To achieve the latter the individual reservoirs must operate at different temperatures due to the largely different vapor pressures\cite{Stan2005}. 

\begin{figure}
\centering
\includegraphics[width=0.45\textwidth]{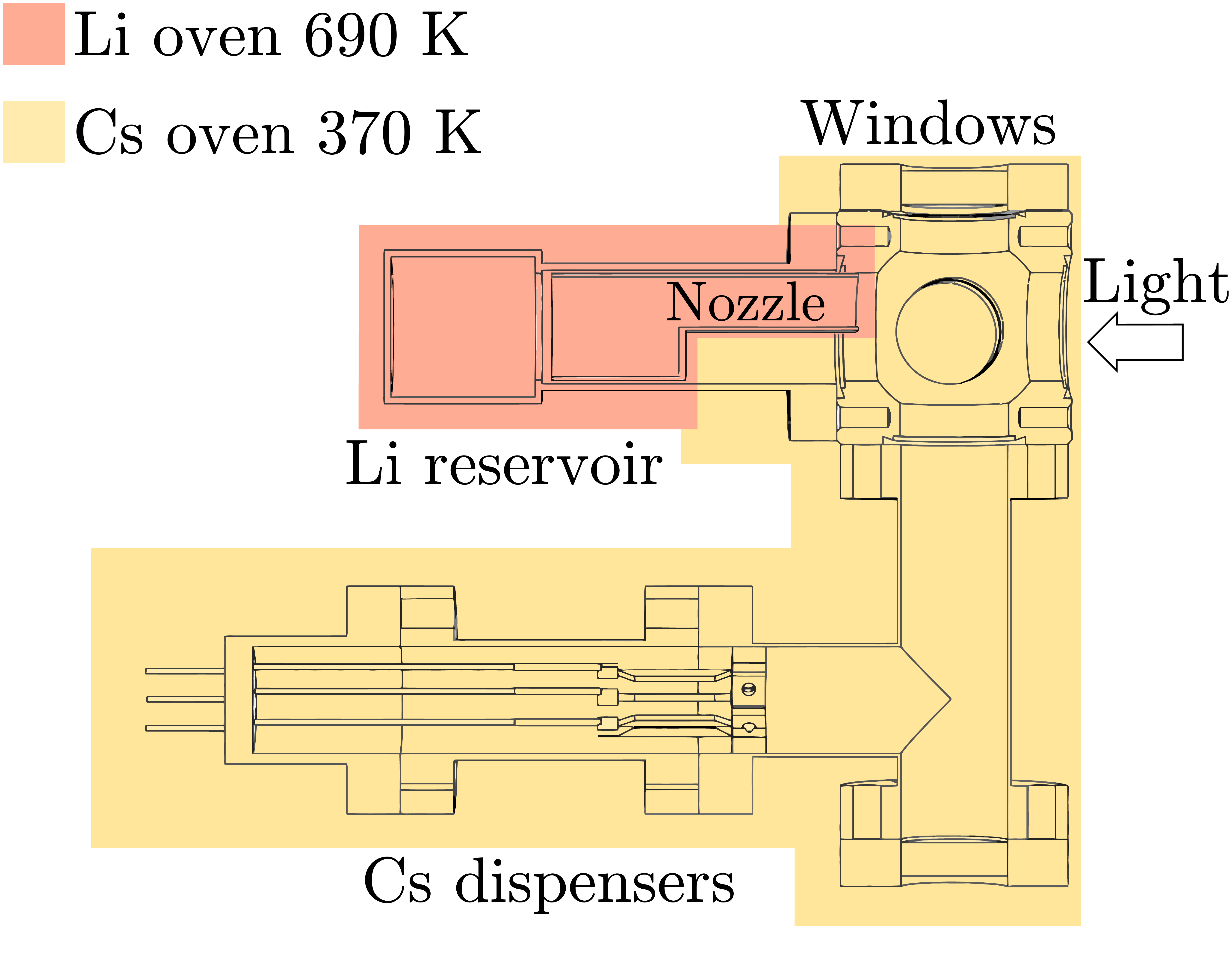}
\caption{\label{oven_schem} Cross-section diagram of dual-species oven. Regions of the oven heated to different temperatures are indicated by shaded areas.}
\end{figure}

Our dual-species oven combines a heated reservoir containing pure chunks of lithium and a dispenser-fed oven for cesium. An oven heated to 690~K by an external heating wire \footnote{ISOMIL-H mineral insulated heating cable} contains approximately 800~mg of 99.9\% pure lithium-6 metal. At this temperature a vapor pressure of $10^{-4}$~mbar is expected \cite{Lide2013}. A nozzle with a semi-circular cross section connects the lithium to the cesium oven. This aids against flow of cesium back into the lithium oven and against one gas displacing the other in the region with line-of-sight through the Zeeman slower to the MOT. The shape of the nozzle means that half of the circular oven aperture gives lithium line-of-sight through the slower and the other half cesium. The nozzle has good thermal contact with the lithium oven to prevent condensation of lithium on its inside surface causing clogging. This setup is displayed in Figure \ref{oven_schem}.

Cesium has a relatively high vapor pressure of $10^{-6}$~mbar at room temperature \cite{Lide2013}. For this reason dispensers containing a cesium salt are employed, instead of using an oven containing pure cesium. This design allows the addition of new cesium atoms to the chamber to be halted almost instantaneously by switching off the electrical current in the dispensers. As a result, the cesium pressure can be more controllably managed and the source is not depleted when not in use. In addition, this is expected to lengthen the lifespan of the ion pump which can suffer from corrosion due to the cesium. Eight dispensers are mounted on a ceramic ring and connected to a power supply via a nickel-wire vacuum feedthrough. The vacuum tube surrounding the dispensers is heated to a temperature of 370~K to avoid condensation of cesium.


In summary, the resultant dual-species oven provides a controllable source of lithium and cesium atoms despite their large difference in vapor pressure. The flux of each species can be independently managed by altering the temperature of different parts of the oven. By using a dispenser source, introduction of cesium can be turned off completely. In addition, using dispensers avoids reactions of cesium with water which greatly simplifies the installation of the oven. Furthermore, this is accomplished in a simple design with differential pumping which ensures minimal impact on the pressure in the MOT chamber.

\subsection{Zeeman slower\label{Zeeman slower}} 

The slowing light is an important ingredient for the implementation of a Zeeman slower. The atomic transitions we use for slowing are $F=4 \rightarrow F'=5$ and $F=3/2 \rightarrow F'=5/2$ of the $D_2$ line for cesium and lithium respectively. In addition, it is necessary to repump the atoms to form a closed cycle, this we do on $F=3 \rightarrow F'=3$ for cesium and $F=1/2 \rightarrow F'=3/2$ for lithium. Both slower and repumper beams have the same red detuning from their respective transitions. The detuning at which we operate the slower is 95~MHz for cesium and 120~MHz for lithium.  

The cesium and lithium slowing light is combined to a single beam on a dichroic mirror and pointed through a window on the axis of the slower, as depicted in Figure~\ref{apparatus_profile}. This beam is slightly convergent such that its intensity at the MOT position is reduced, but is sufficiently high in the slower.

The Zeeman slower itself comprises nine coils on a 28~mm diameter copper tube. Copper plates of 1~mm thickness separate the coils and act as heatsinks. The coils consist of 28, 19, 19, 17, 15, 14, 11, 9 and 6 layers respectively, with 13 windings in each layer. The coils are wound using Kapton-insulated copper wire of 1.72~mm high and 3.14~mm wide rectangular cross section. The rectangular cross section allows uniform flat layers to be wound. 

\begin{figure}
\includegraphics[width=0.45\textwidth]{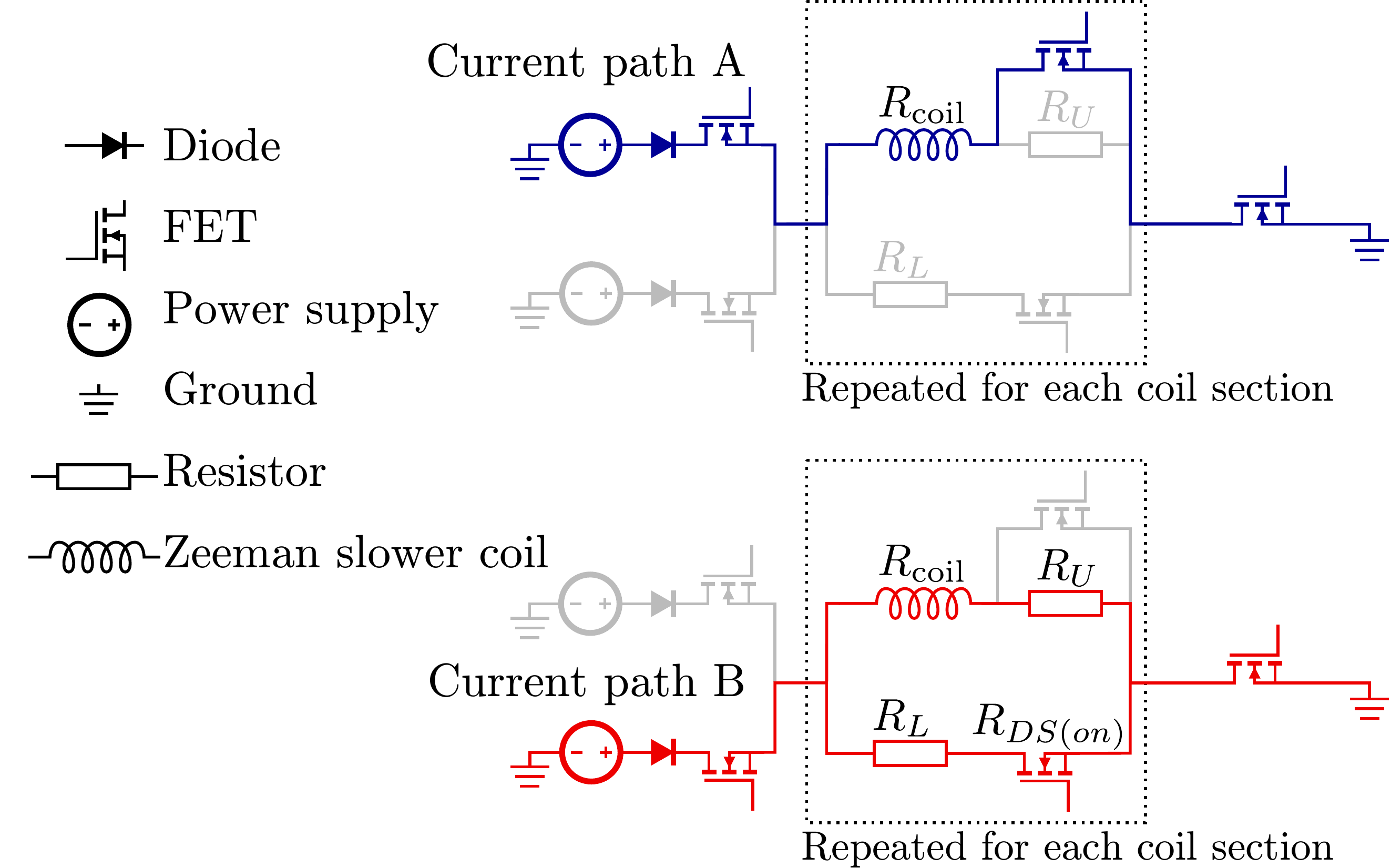}
\caption{\label{electronics_schem} Schematic of electronics. Current path A is active when the slower is used for lithium. Current path B shows operation for cesium.}
\end{figure}

To realize the designed field profiles as shown in Figure~\ref{calc_profiles_design} for lithium, a constant current of 11~A is needed throughout the coils whereas the cesium profile requires a different current in each coil with much lower maximum current of 2.0~A. An electronic circuit was introduced to switch between the two magnetic field profiles. A schematic of the main elements of the circuit is shown in Figure~\ref{electronics_schem}. Separate power supplies are used for each of the two current settings. Two field effect transistors (FET) with a differential FET driver are used to switch between them. Equivalently, this could be achieved using a single controllable supply. Another FET in series with the circuit acts as an on/off switch for the entire field. The current in each coil of the slower is tailored by introducing an alternative path for the current so that only a chosen fraction goes through the coil. This current shunt was implemented with differentially driven FETs. Resistor values were chosen to achieve the design current in each coil with minimum power dissipation. Figure~\ref{electronics_schem} shows the current path when operated for lithium (path A) and for cesium (path B).

The designed system has two modes of operation. Current path A is produced by having the upper FETs on while the lower ones are off. All of the current coming into the coil section passes through the coil and all of the resistors are bypassed to reduce power dissipation. Path B occurs when only the lower FETs are on. In this case the current is split between the upper and lower path according to
\begin{equation}
{{I_U} \over {I_L}} = \frac{R_L+R_{DS(on)}}{R_U+R_{\text{coil}}},
\end{equation}
where $R_{DS(on)}$ is the drain-to-source resistance of the FET when on, $R_{\text{coil}}$ it the resistance of the coil, and $R_U$ and $R_L$ are indicated in Figure~\ref{electronics_schem}.

\begin{figure}
\includegraphics[width=0.45\textwidth]{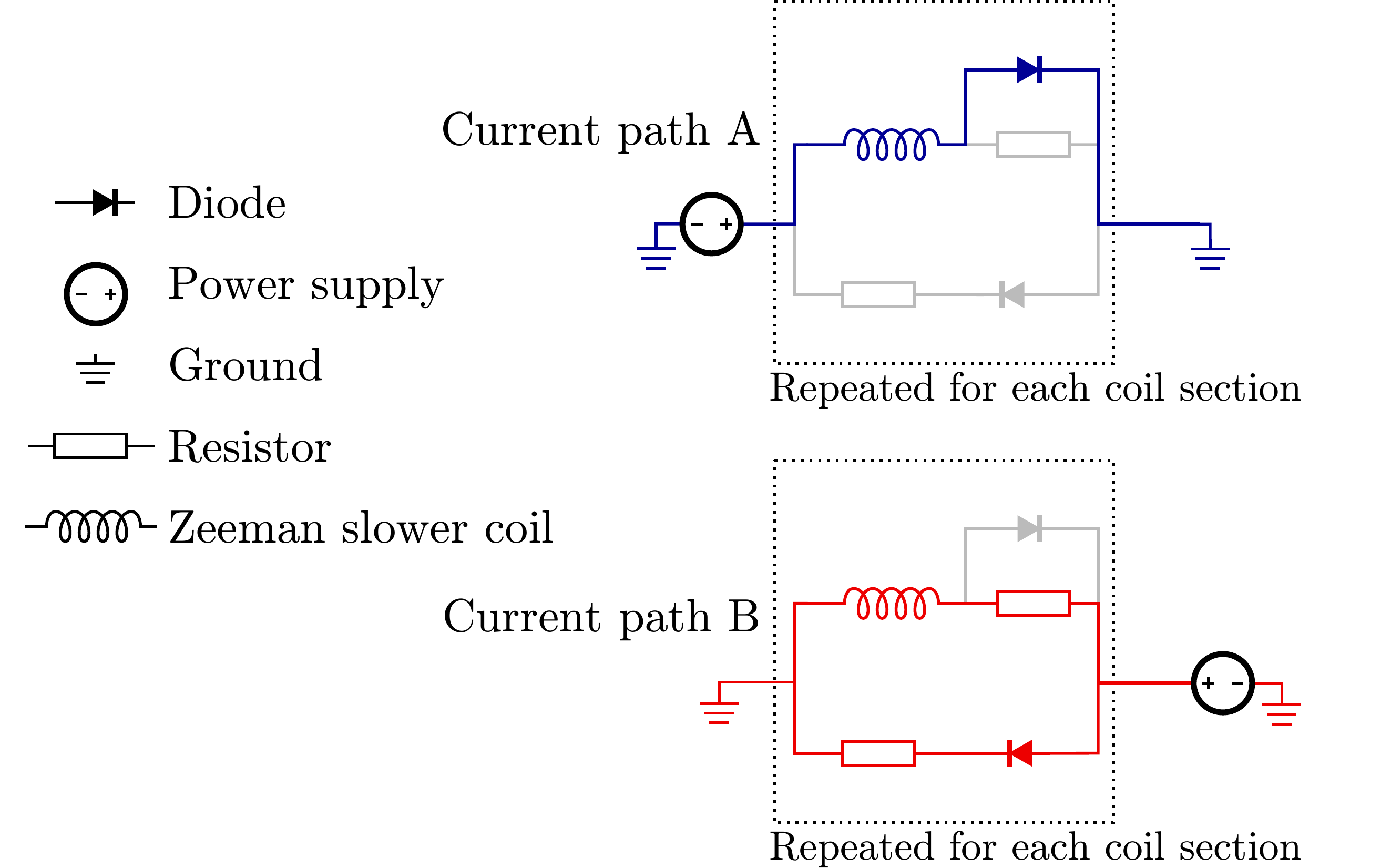}%
\caption{\label{diode_circuit} Circuit schematic for each coil section where the current direction is switched. Current path A is active when the slower is used for lithium. Current path B shows operation for cesium.}
\end{figure}

Our design can be used to switch between any type of Zeeman slower. For switching between a decreasing or increasing-field slower to a spin-flip slower the coil sections, where  the polarity is inverted, should be within a standard H-bridge circuit. For coil section circuits with inverted current the FETs cannot be used as the source-to-drain voltage must be positive. Instead, they can be replaced with low voltage drop diodes, as shown in Figure~\ref{diode_circuit}. In this design, the polarity switch is advantageous as it is used to open the shunt current path using the diode.

To minimize power dissipation, FETs with small $R_{DS(on)}$  and low voltage drop diodes are used. For our system the total maximum power dissipation of the electronics is only 30~W. This figure represents a relatively small increase on the 100~W dissipated by the slower coils themselves.

The presented scheme is versatile and can easily be adapted to a wide variety of scenarios. A voltage controlled current limiter circuit using a FET could replace each resistor. This would allow full control of the magnetic field profile from a control system. This could be used to perform automated optimization or to switch between any number of field profiles and use the system for more than two atomic species.

\section{Measurements and Discussion\label{Measurements and Discussion}}

We optimized the performance of the slower by maximizing the initial atom loading rates of the MOT. For each species, the MOT consists of three retro-reflected beams along orthogonal axes with a $1/{e^2}$ diameter of 24~mm. The two MOTs are positioned a few millimeters apart from each other\footnote{A small displacement of the central position of the two MOTs avoids interspecies collisional losses.} in the MOT chamber and can be operated at the same time, but are loaded sequentially. The cesium MOT uses approximately 24~mW of cooling light and 1.6~mW of repumping light in each beam. The lithium MOT uses 28~mW of cooling light and 24~mW of repumper in each beam. Anti-Helmholtz coils create magnetic field gradients of 14~G/cm and 20~G/cm for the cesium and lithium MOTs respectively. We can load MOTs with a total number of $6\times 10^8$ lithium atoms and $7\times 10^7$ cesium atoms.

\begin{figure} \centering
\begin{tikzpicture}[      
        every node/.style={anchor=south west,inner sep=0pt},
        x=1mm, y=1mm,
      ]   
     \node (fig1) at (0,0)
       {\includegraphics[width=0.45\textwidth]{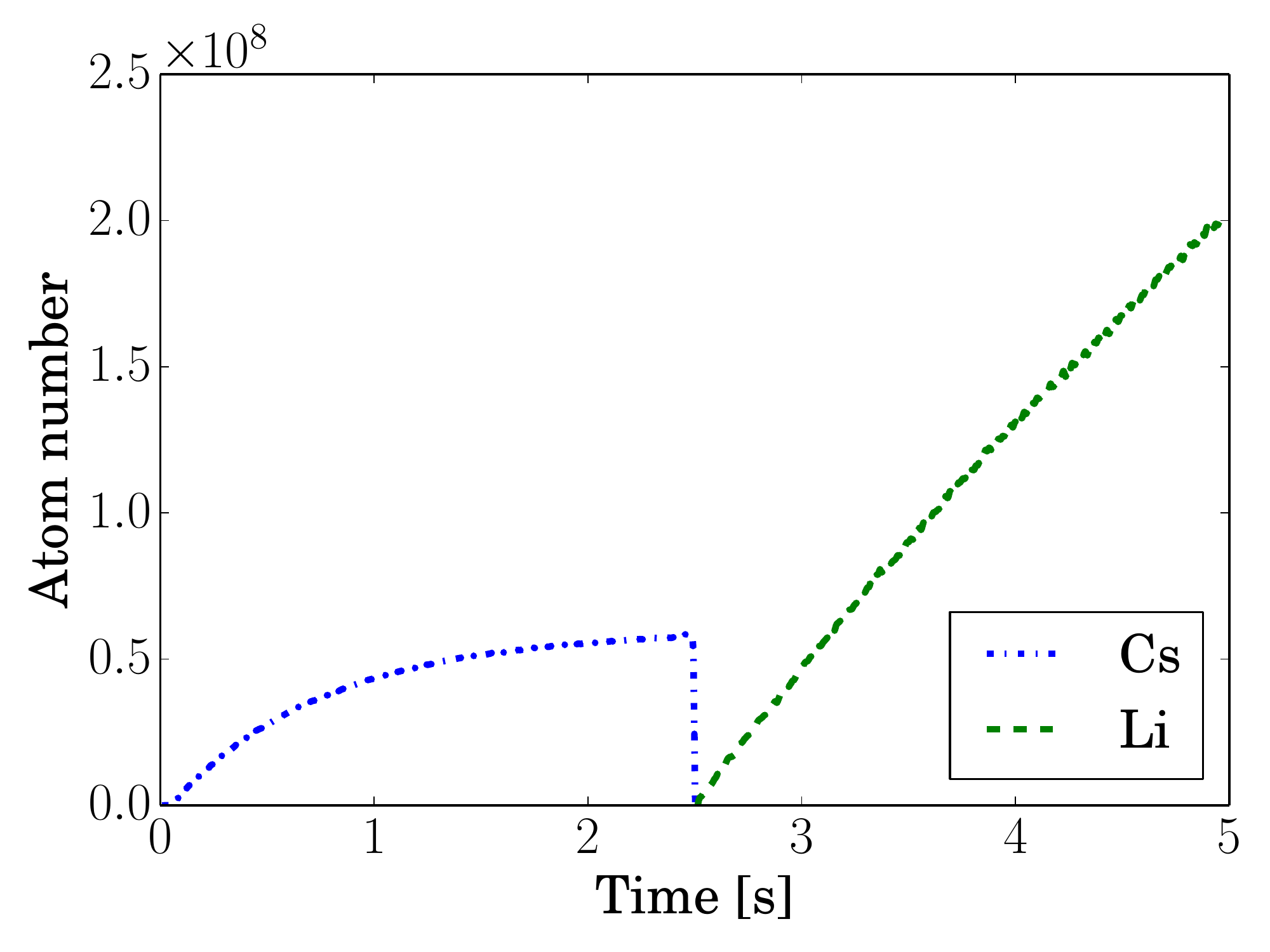}};
     \node (fig2) at (14,25)
       {\includegraphics[width=0.22\textwidth]{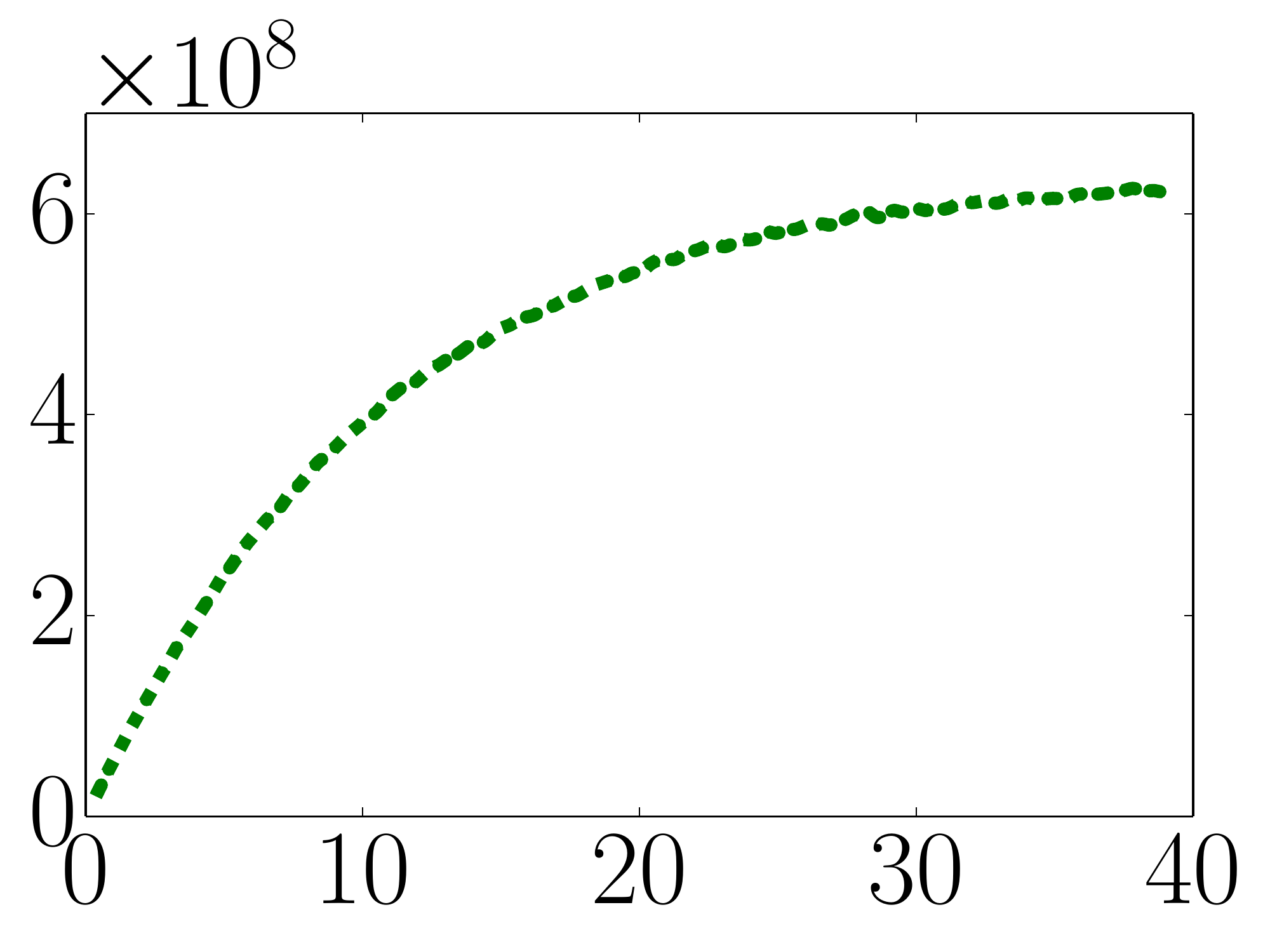}};  
\end{tikzpicture}
\caption{\label{li_cs_switch} Atom number from calibrated photodiode signal in the lithium and cesium MOTs during loading. Between 0 and 2.5~s the cesium MOT is loaded. Between 2.5 and 5~s a lithium MOT is loaded. The curves shown are the average of 6 measured traces. The inset shows the lithium loading curve to saturation of the MOT. }
\end{figure}

Figure~\ref{li_cs_switch} shows loading a cesium MOT, immediately followed by loading a lithium MOT at 2.5~s. This measurement demonstrates the system operating for both atom species as well as the short switching time of the slower. At 2.5~seconds the slower magnetic field profile and MOT coil currents are switched to the optimal setting for loading a lithium MOT. Also at this time the cesium slower light and MOT beams are switched off and the lithium slower light and MOT beams switched on.

We optimized the performance of the slower by maximizing the loading rates of the MOTs via fine tuning of the magnetic slower field profile and varying the amount of used slowing light.
A measurement of fluorescence using a photodiode shows the exponential growth of the number of atoms in the MOT after opening the atomic beam shutter, see Figure \ref{li_cs_switch}. The photodiode voltage was calibrated to atom number using absorption images of the atom cloud and the initial loading rate was found by using exponential fits to the loading curves.

\begin{figure}
\includegraphics[width=0.45\textwidth]{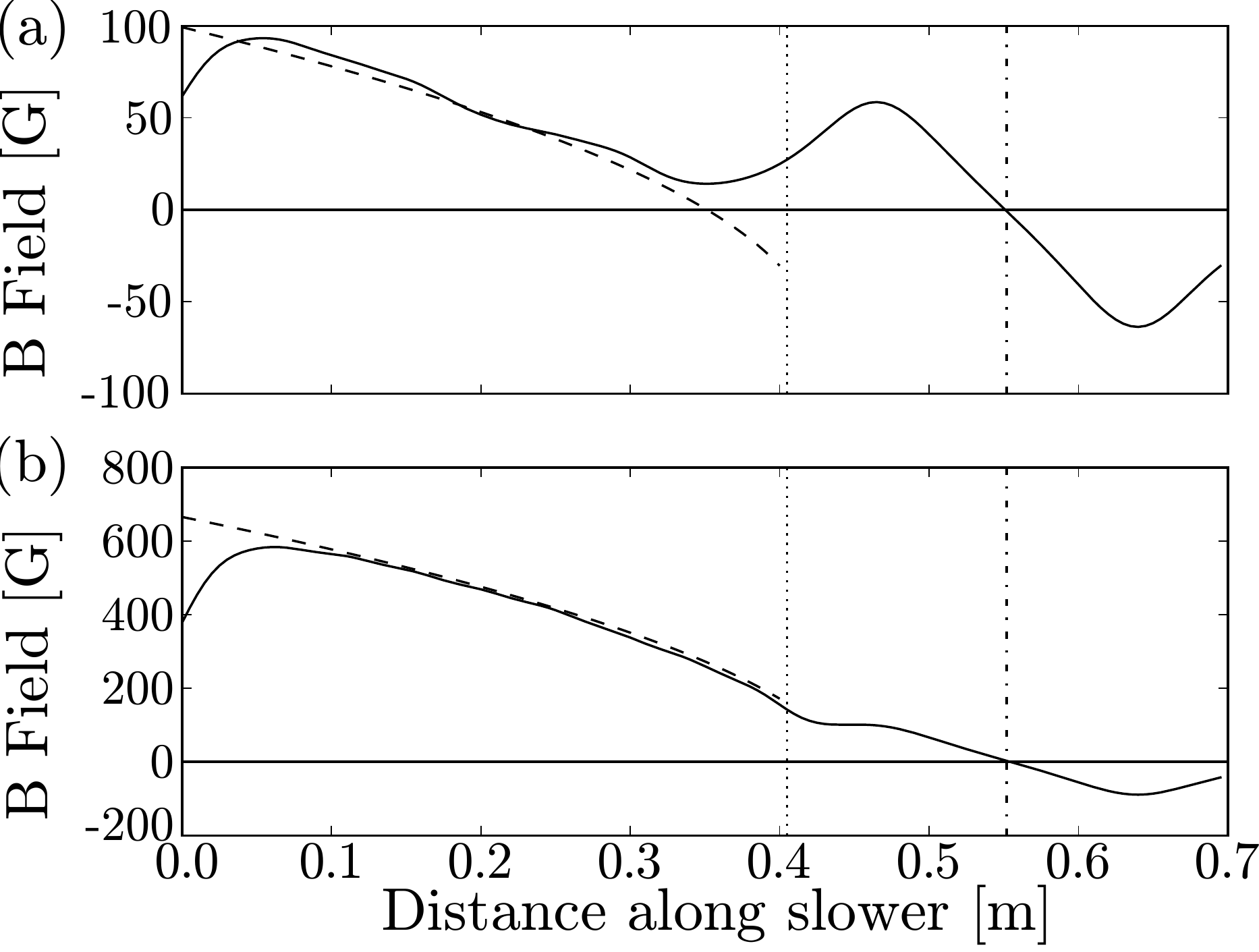}%
\caption{Calculated magnetic field profiles of the Zeeman slower for (a) cesium and (b) lithium. Solid lines show the expected magnetic field magnitude using measurements of the current in each coil section of the optimized slower. Dashed lines are the expected ideal magnetic field profile to achieve constant deceleration, using $\eta=0.3$ for lithium and $\eta=0.8$ for cesium. The vertical dotted line indicates the end of the slower coils at 0.405~m. The vertical dash-dotted line indicates the central position of the MOT.}
\label{calc_profiles_actual}
\end{figure}

For the initial design $\eta=0.5$ was assumed for both species. However, having the ability to tailor the field, one can optimize the MOT loading rate for the actual, achieved deceleration. In order to find the magnetic field profile which maximizes the loading rate of the MOT, values of resistors were changed in the Zeeman slower. For cesium we find that a field profile matching $\eta=0.8$ resulted in the largest MOT loading rate. For lithium operating the slower at 8.2~A proved to provide the highest MOT loading rate. This current matches an expected ideal field with $\eta=0.3$, as displayed in Figure~\ref{calc_profiles_actual}~(b). The expected magnetic field profiles with $\eta=0.3$ (Li) and $\eta=0.8$ (Cs) and the actual field produced by the experimentally optimized currents are shown in Figure~\ref{calc_profiles_actual}. During optimization, we found that a decreasing-field slower resulted in a higher loading rate of the cesium MOT. The significant change from designed to optimized field profiles highlights the advantage of a design which permits optimization.

\begin{figure}
\includegraphics[width=0.45\textwidth]{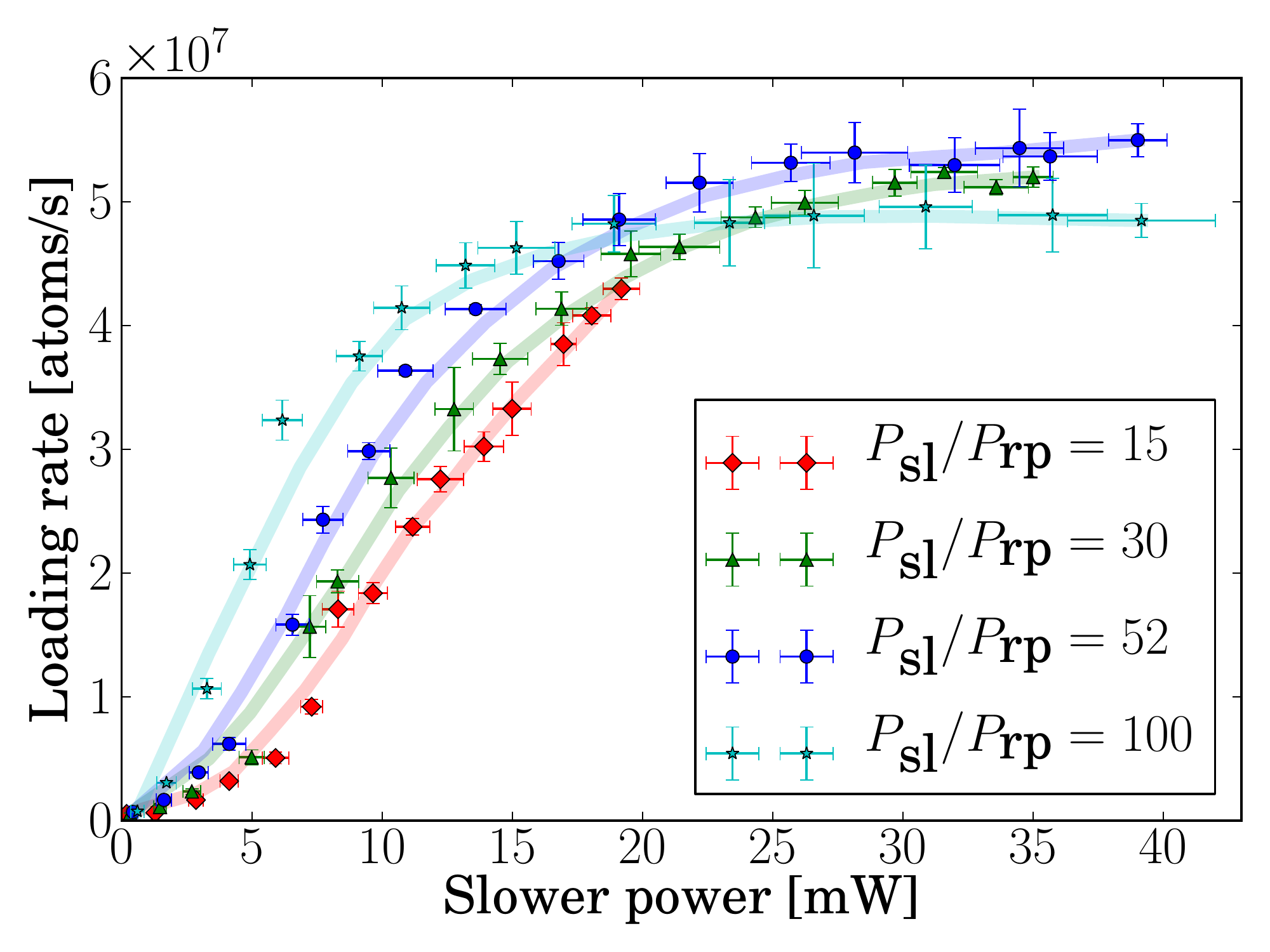}
\caption{\label{optimize} Slower light power optimization for cesium. Each set of points displays an average of three measurements of the MOT loading rate for a different ratio of slower to repumper light. Lines show moving averages to guide the eye.}
\end{figure}

\begin{figure}
\includegraphics[width=0.45\textwidth]{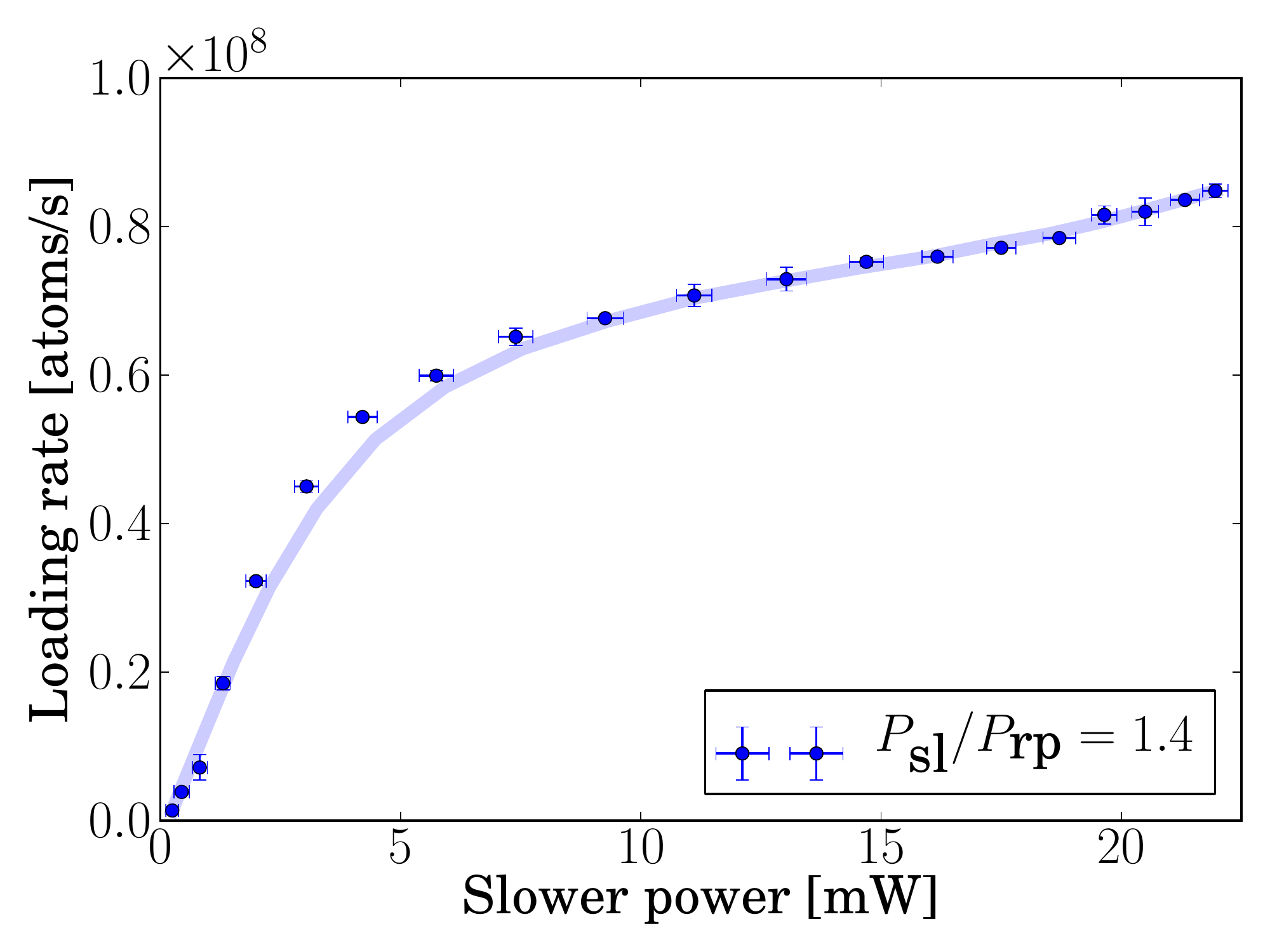}
\caption{\label{li_optimize} Slower light power optimization for lithium for a slower to repumper power ratio of 1.4. Points are an average of two measurements. The line shows a moving average to guide the eye.}
\end{figure}

Additionally the amount of cooling and repumper light in the Zeeman slower was adjusted for optimum MOT loading rates. For technical reasons of the cesium optical setup, the most straightforward manner of investigating this was to choose a ratio of slower to repumper and then vary the total power. This was repeated for four slower to repumper ratios. The resulting curves for cesium MOT loading rate versus power are shown in Figure~\ref{optimize}. Using 30~mW of cooler and 0.6~mW of repumper in the convergent 50~mm diameter beam gave the largest loading rate.
A similar optimization process was carried out for lithium. A plot of loading rate versus total slower light power is shown in Figure~\ref{li_optimize}. With a ratio of slower to repumper light of $1.4$ a fast loading rate of approximately $1\times10^8$~atoms/s is achieved \cite{Marti2010, Taglieber2006, Schloder1999}.

\begin{figure}
\includegraphics[width=0.45\textwidth]{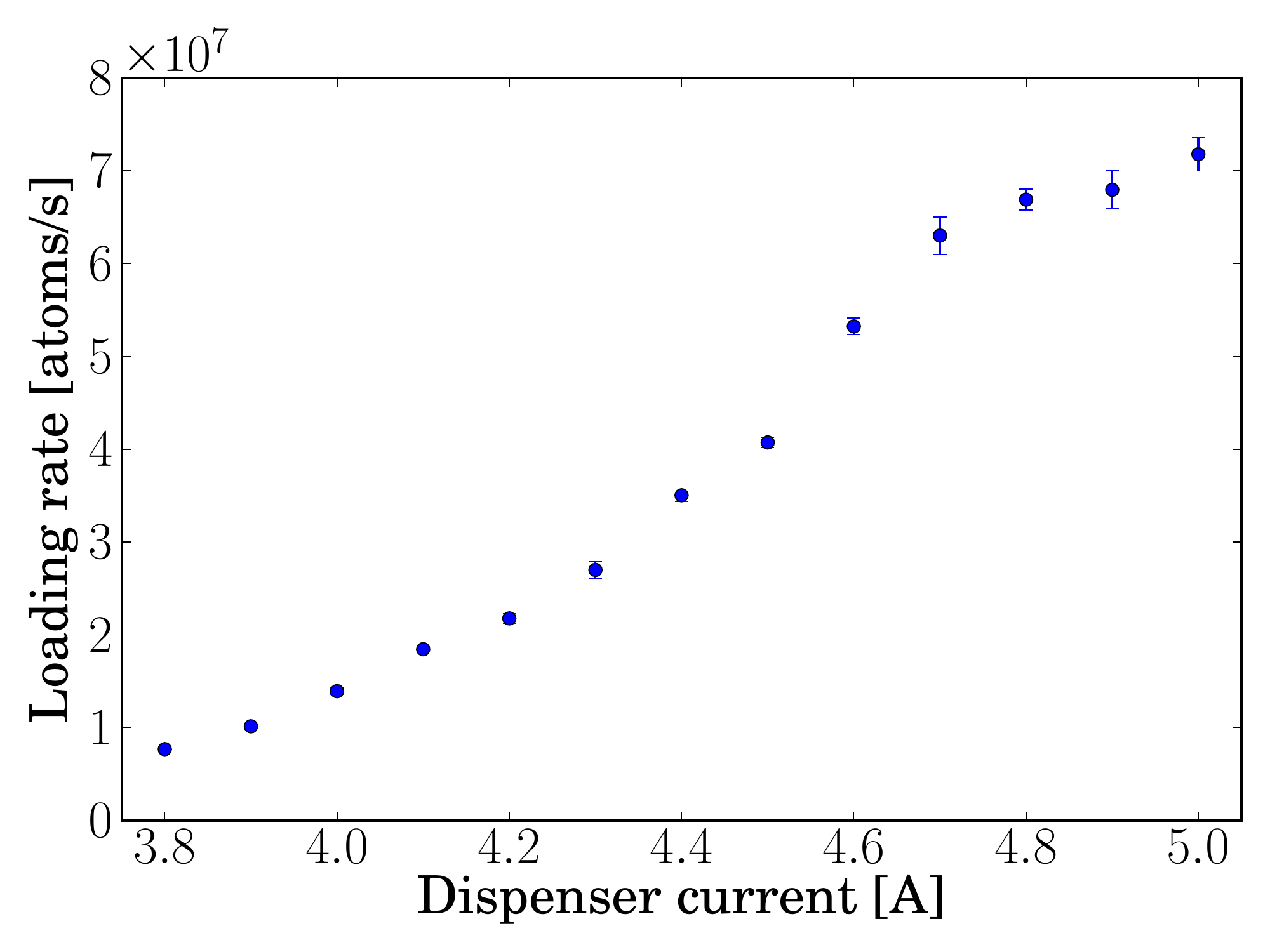}
\caption{\label{dispenser current} Cs MOT loading rate dependence on the current using in dispensers in the oven. Points are an average of ten measurements.}
\end{figure}

For cesium the atomic beam flux and thus the loading rate of the MOT also depends on the current in the dispensers, as shown in Figure~\ref{dispenser current}. We use four out of eight dispensers in the oven at a time with a current of 4.8~A and achieve a good MOT loading rate\cite{Kraft2006} for cesium of $7\times10^7$~atoms/s. This demonstrates that the dispenser source can perform as well as a conventional effusive oven as an atom source for a Zeeman slowed beam, whilst retaining the advantages mentioned previously in Section~\ref{oven}.

\section{\label{Conclusion}Conclusion}

We have presented a dual-species atom source consisting of a single oven and a single Zeeman slower with an electronically switchable field profile. Details of the electronic and mechanical implementation of the slower in addition to the optimization of both the magnetic field profile and laser powers were given. This design is both simple and versatile and may be easily implemented in new experiments or with existing slowers originally designed for a single atomic species.

\section{\label{Acknowledgments}Acknowledgments}
We would like to thank R. Kollengode Easwaran, IIT Patna for his contribution at an early stage of the experiment and C. Koller for careful reading of the manuscript. A.P-M. acknowledges CONACYT, SEP and the Mexican Government for their financial support. J.W. acknowledges the
support from the Royal Society K. C. Wong Postdoctoral Fellowship. This project was supported by EPSRC grant EP/K023624/1 and by the European Commission grant QuILMI - Quantum Integrated Light Matter Interface (No 295293).

\bibliography{library}

\end{document}